\documentclass[prl,showpacs,twocolumn,preprintnumbers,amsmath,amssymb]{revtex4}

\usepackage{graphicx}
\usepackage{dcolumn}
\usepackage{bm}
\usepackage{psfrag}

\newcommand{\beq}{\begin{equation}}
\newcommand{\eeq}{\end{equation}}
\newcommand{\bea}{\begin{eqnarray}}
\newcommand{\eea}{\end{eqnarray}}

\newcommand{\bR}{{\bm{R}}}

\newcommand{\ome}{\omega}

\begin{document}


\title{ Nonequilibrium noise in electrophoresis: the microion wind}

\author{Suropriya Saha}
\email{suropriya@physics.iisc.ernet.in}
\author{Sriram Ramaswamy}%
 \email{sriram@physics.iisc.ernet.in}
\affiliation{%
Indian Institute of Science
}%

\date{\today}

\begin{abstract}
We analyze theoretically the dynamics of a single colloidal particle in
an externally applied electric field. The thermal motions of microions
lead to an anisotropic, nonequilibrium source of noise, proportional to
the field, in the effective Langevin equation for the colloid. The
fluctuation-dissipation ratio depends strongly on frequency, and the
colloid if displaced from its steady-state position relaxes with a velocity
not proportional to the gradient of the logarithm of the steady-state
probability.

\end{abstract}

\maketitle

Colloids in electric fields present challenging problems in nonequilibrium
statistical mechanics, as a result of the interplay of hydrodynamic and
electrostatic interactions \cite{ristenpart}, \cite{sides}. Electrophoresis and
aggregation into chains, crystalline phases and fractals are just some of the
diverse phenomena seen in these systems \cite{trau}, \cite{yeh}, \cite{sood}.
However, a thorough theoretical understanding of these effects must begin with
a single colloidal particle in an electric field. 
Important progress in this direction has been made by Squires \cite{squiresM}
who has shown that particles stably suspended by the interplay of external
fields, boundaries and hydrodynamic flow at low Reynolds number respond to
disturbances as though they were in equilibrium in an effective potential.  The
effective particle velocity divided by the Stokes drag is treated as a
pseudoforce which subsumes all hydrodynamic interactions of the colloidal
particle and surrounding counterions  with the wall. This pseudoforce, it turns
out, can be written as the gradient of a scalar `pseudopotential'. A major
success of this approach was the explanation of apparent like-charge attraction
in certain experiments \cite{squires2}, \cite{Larsen}. 

To extend this formalism beyond the average motion of the colloid, note that it
is subject to two kinds of stochastic forces. One, which is taken into account 
in \cite{squiresM}, is thermal noise corresponding to
viscous dissipation, with variance proportional to temperature times Stokes
drag. The other enters as follows: the local charge density $\rho$ fluctuates
because of thermal motions of the microions; in the presence of an
imposed electric field ${\bf E}$, this means the electric force density $\rho {\bf E}$ in
the Stokes equation fluctuates as well. This results in a fluctuating
contribution to the motion of the colloid, with variance proportional to
temperature times the square of the electric field and correlations controlled by microion motion
\cite{sriprost}. This nonequilibrium coloured noise, ignored so far as we know in
earlier work \cite{squiresM}, is the main subject of this Letter.

We summarize our main results before presenting details of the work. We have
studied two systems: (I) a neutrally buoyant colloid drifting uniformly under
an electric field in an unbounded fluid; and (II) a colloid with density higher
than the fluid, stably levitated by the balance between gravity sedimenting it
towards a wall and an electric field driving it away. In I the excess noise
variance is $1-10^3$ times the thermal noise variance; in II it is found to be
at least as large as the thermal noise, and multiplicative in nature [see Fig.1
(a)]. In both cases the extra noise is anisotropic -- nearly an order of
magnitude stronger along $\bm{E}$ than transverse to it. If $S_\omega$ and
$\chi_\omega$ are the correlation and response functions of the colloid
position at frequency $\omega$, we see that $ \omega S_\omega/2 \mbox{Im}
\chi_\omega \equiv T_{\omega}$, which should reduce to the thermodynamic
temperature at thermal equilibrium, instead changes by a factor of $2$ as
$\omega$ changes from $0$ to $5.5 D \kappa^2$ ($D=$ typical diffusivity of the
counterions and impurity ions hereafter collectively called microions,
$\kappa^{-1}= $ Debye screening length), Fig 1(b). 
\begin{figure}[h]
\begin{center}
\psfrag{a}{$\kappa a = 1$}
\psfrag{b}{$\kappa a = 0.3$}
\psfrag{c}{$\kappa a = 3$}
\psfrag{d}{$\kappa a = 5$}
\psfrag{r}{$R$}
\psfrag{Ratio}{$Ratio$}
\psfrag{T}{$T_{\omega}$}
\psfrag{w}{$\omega$}
\psfrag{e}{Sys I}
\psfrag{f}{Sys II}
\psfrag{lambdac}{\Large $\lambda_{x-y}$}

                \includegraphics[angle=0,width=2.2in]{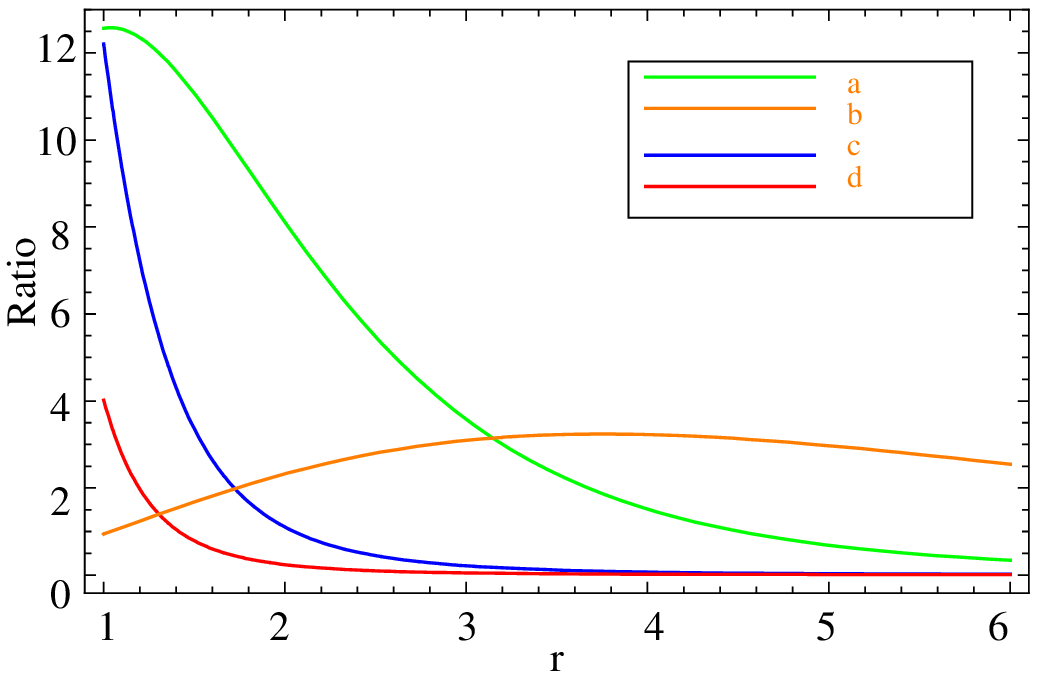}
                \includegraphics[angle=0,width=2.2in]{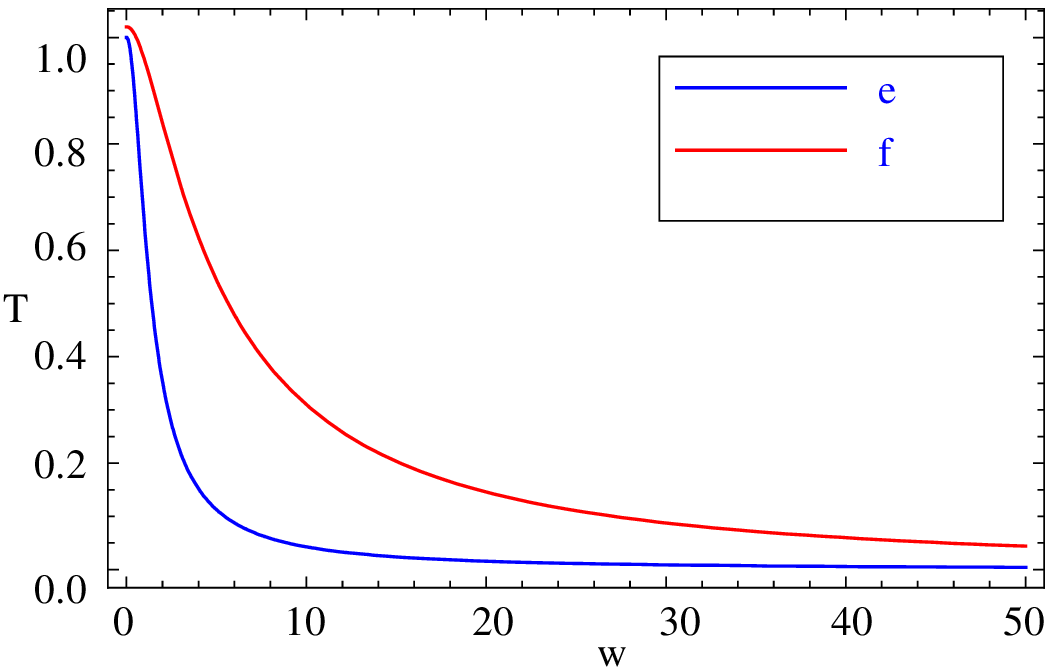}
                
                \caption{
                         (a) Plots showing the ratio of excess noise to thermal noise strength as a function of the distance from the wall $z$ for different screening lengths.\\
                         (b) Plots showing $T_{\omega}$ vs. $\omega$ for I and II where $\omega$ has been scaled by $D \kappa^2$.}
        \end{center}
\end{figure}


We also find the steady-state solution $P_\infty(\bm{R})$ to the Fokker-Planck
(FP) equation for the probability density of the colloid position $\bm{R}$. We
find that the effective potential $U_{eff}(\bm{R}) \equiv -k_B T \ln P_\infty
(\bm{R})$, where $T$ is the thermodynamic temperature, is shallower than
Squires's effective potential for the same problem, Fig. 2. The excess noise
causes larger excursions than would arise from the bare thermal noise. 
Thus departures from equilibrium behaviour should be evident even in single-particle 
experiments.
\begin{figure}[h]
\begin{center}
\psfrag{a}{$U_{eff}$}
\psfrag{b}{$U_{eff}^{0}$}
\psfrag{c}{$U_{eff}$}
\psfrag{d}{$U_{eff}^0$}
\psfrag{r}{$R$}
\psfrag{P}{$U_{eff}$, $U_{eff^0}$}
\psfrag{u}{$U_{eff}$, $U_{eff}^0$}
                \includegraphics*[angle=0,width=2.2in]{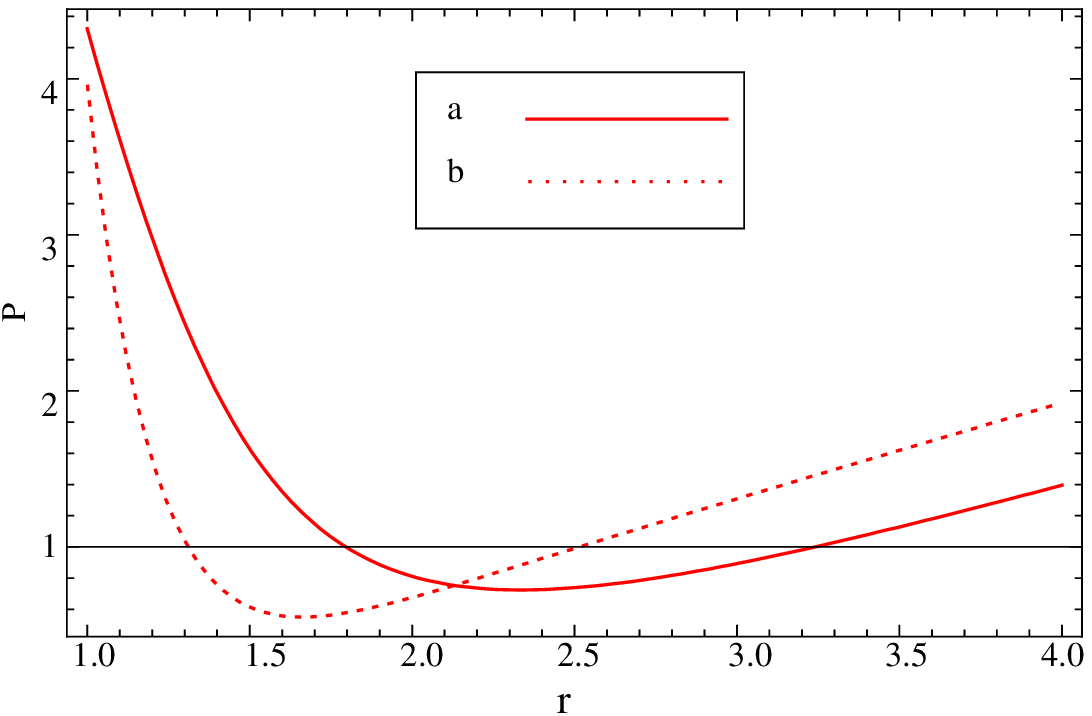}
                \includegraphics*[angle=0,width=2.2in]{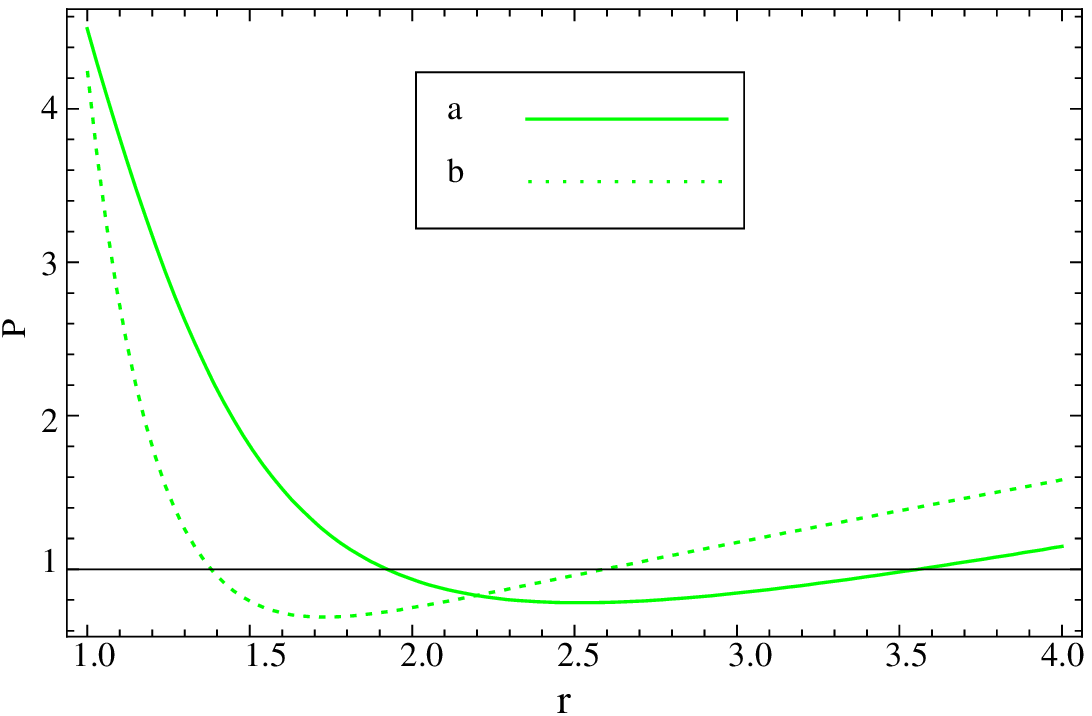}
                               \caption{ The effective potential as computed from the Fokker Planck equation including the excess noise (solid line), compared to that in \cite{squiresM} (dotted line) for $\kappa a = 4$, $a=1 \mu$m, $\phi_0 = 4$ V for (a) iron core (red), (b) silica core.}
        \end{center}
\end{figure}
We now show how these results were obtained, beginning with case I. The colloid consists of a charge $Q$, surrounded by a counterion cloud of total charge $-Q$. The counterions plus ionic impurities, for short the microions, are assumed for simplicity to have identical mobility $\mu$ and unit valency. For simplicity, we ignore advection of the microions by the velocity field. Thus the continuity equations for the densities $n^+$ and $n^-$ of positive and negative microions read
\begin{equation}
        \frac{\partial{n^+}}{\partial{t}} = D\nabla^2n^{+} - \mu\nabla\cdot(n^{+}\bm{E})+\bm{\nabla} \cdot (\sqrt{2 n^+ D}\bm{f^+}),
\label{poscont}
\end{equation}

\begin{equation}
        \frac{\partial{n^-}}{\partial{t}} = D\nabla^2n^{-} + \mu\nabla\cdot(n^{-}\bm{E})+\bm{\nabla} \cdot (\sqrt{2 n^- D}\bm{f^-}),
\label{negcont}
\end{equation}
where $\bm{f^+}$ and $\bm{f^-}$ are unit strength, independent white Gaussian noises, 
$D = k_B T \mu$ is the ionic diffusivity, $\mu$ being the Ohmic mobility, and $\bm{E}$ is the total electric field.

Momentum conservation at zero Reynolds number, as appropriate for colloids, with thermal noise, leads to the fluctuating Stokes equation for the hydrodynamic velocity field $\bm{v}$ of the incompressible suspension:
\begin{equation}
        \eta \nabla^2 \bm{v} - \bm{\nabla} p + \rho \bm{E} +\sqrt{2k_{B}T\eta}\bm{\xi} = 0; \bm{\nabla} \cdot \bm{v} = 0,
\label{momcons}
\end{equation}
where $\bm{\xi}$ is a unit strength, divergence free conserving Gaussian white noise.
We close the system with the Poisson equation
\begin{equation}
\bm{\nabla} \cdot \bm{E}  = \frac{\rho}{\epsilon_{0}},
\end{equation}
where the total charge density $\rho =\rho^{Q}+\rho^{mi}$, $\rho^{Q} \equiv Q \delta (\bm{r} - \bm{R}(t) )$ and $\rho^{mi} \equiv e(n^+-n^-)$ being the colloid and microion charge densities respectively. The colloid is approximated as a point charge $Q$, and the effects due to nonzero size of the particle are taken care of by introducing an ultraviolet cutoff  $2 \pi/a$ in Fourier space.
We assume the colloid position $\bm{R}(t)$ simply moves with the fluid,
\begin{equation}
        \frac{\partial \bm{R}(t)}{\partial t} = \bm{v}(\bR(t)).
\label{colloid}
\end{equation}

The boundary conditions to be satisfied are: $\bm{E}(\bm{r}\rightarrow \infty)
= E_0 \bm{\hat{z}}$ where $E_0$ is the imposed field, and $n^\pm(r \rightarrow
\infty) = n_0$, the mean concentration of either species of ions. To make the
problem tractable we linearize the equations. The electrostatic force density
in the Stokes equation thus becomes $\rho^{0} E_{0} \hat{\bm{z}}$ where
$\rho^{0}$ is the charge density in the absence of the electric field. 
The charge density has a
steady and a fluctuating part: $\rho^{0}(\bm{r},t) = \bar{\rho}(\bm{r})+
\delta \rho (\bm{r},t)$. Solving (\ref{poscont},\ref{negcont}) for the charge
densities in the absence of a field we get $\delta \rho(\bm{r},t)
\propto-i(\bm{q}.\bm{f}^{+}_{\bm{q}\omega} - \bm{q}.\bm{f}^{-}_{\bm{q}\omega}
)/(q^{2} - i\frac{\omega}{D} + (\kappa a)^{2})$. The term $\delta\rho E_{0}
\hat{\bm{z}}$ becomes an additive nonequilibrium Gaussian noise in the Stokes
equation and thus contributes a noise proportional to $E_{0}$ in the equation
of motion for the colloid. Inserting the solution into the Stokes equation and
evaluating $\bm{v}(\bR(t))$, we find the effective Langevin equation 

\begin{equation}
\frac{d \bm{R}}{dt} = \bm{v}_0 + \bm{f} + \bm{\zeta}
\end{equation}
for the colloid, where $\bm{f}$ is the thermal Gaussian white noise with variance $2 k_B T/ \Gamma$, $\Gamma \propto \eta a$ is the Stokes drag coefficient, $\zeta$ with variance $\propto E_0^2$ is the nonequilibrium excess noise and $\bm{v}_0 = \alpha Q E_0 \kappa^{-1}/\eta a^2$ is the steady electrophoretic velocity, $\alpha$ being a numerical factor of order unity. 
We measure the relative zero frequency strengths of excess and thermal noise by
\begin{equation}
 \mathcal{R}_{ij} \equiv \frac{\int_0 ^\infty \mbox{d}t\langle \zeta_i (0) \zeta_j (t)\rangle}{\frac{1}{3}\int_0 ^\infty \mbox{d}t\langle f_k (0) f_k (t)\rangle}
\end{equation}
and find a strong anisotropy in the excess noise and hence in $\mathcal{R}$; $\mathcal{R}_{zz}/\mathcal{R}_{xx} \simeq 8$. Further, to estimate the typical scale of $\mathcal{R}_{ij}$, take  $ (\kappa a )^{2} = 2 a^{2}e^{2}n_0/ \epsilon k_{B}T$, and the ionic mobility $\mu \simeq e/ \eta a_{mic}$ where $a_{mic}$ is the ionic radius. We then find that 

\begin{eqnarray}
\mathcal{R}_{zz} \propto \frac{\epsilon E_{0}^{2} \kappa^{-1} a a_{mic.}}{k_{B}T},
\label{rat}
\end{eqnarray}
which is the ratio of electrostatic to thermal energies in a volume 
involving three different lengths. 
Choosing typical experimental parameters $E_{0}= 10^{5}$ V$/$m, $D=10^{-9} m^2 s^{-1}$, 
$\kappa a =0.3$, $\mathcal{R}_{zz} $ varies from $300$ to $0.5$ for $a$ varying from $10 \mu$m 
to $0.1 \mu$m.


The excess noise should not be thought of as an effective temperature: it is not only 
anisotropic but also strongly frequency dependent. This is clearly seen through the ratio
$T_\omega \equiv \omega S_\omega / 2 \mbox{Im} \chi_\omega$
where $S_\omega$ and $\chi_\omega$ are the correlation and response functions of the colloid position as a function of frequency $\omega$, evaluated as follows. A body force density $[-K \bm{R}(t) + \delta \bm{h}(t)]\delta(\bm{r} - \bm{R}(t))$ is added to the Stokes equation. The term in $K$ holds the co-ordinate $\bm{R}$ at a stationary mean value and $\delta \bm{h}(t)$ is a perturbation which leads to a shift $\delta \bm{R}(t)$ in the particle position. We calculate $\chi_\omega \equiv \langle \delta \bm{R}_\omega \rangle / \delta \bm{h}_\omega$ and $S_\omega \equiv \int \mbox{d}t \mbox{e}^{i \omega t} \langle \delta \bm{R}(0) \delta \bm{R}(t)\rangle$  and find  
\begin{eqnarray}
\frac{T_\omega}{T}
&=& 1 + \lambda \int{\frac{1}{((q^{2} + (\kappa a)^{2})^{2} + \frac{\ome^{2}}{\mbox{D}^{2}})} \mbox{d}^3 q  } 
\nonumber \\
&=& 1 + \lambda \frac{\sqrt{2}(\kappa a)^{3}}{\sqrt{(\kappa a )^{4} + \frac{\ome^{2}}{\mbox{D}^{2}} } \sqrt{\sqrt{(\kappa a )^{4} + \frac{\ome^{2}}{\mbox{D}^{2}} }+ (\kappa a)^{2}}     },
\nonumber
\end{eqnarray}
leading to the results in Fig. 1b.

A more experimentally accessible system is case II, a colloid whose gravitational sedimentation towards a wall is countered by an electric field pushing it away. For this system the boundary conditions are $\phi (z = 0) = \phi_0$, $\phi (z \rightarrow \infty) = 0$, and no slip and no penetration: $\bm{v} (x,y,0) =0$.

The electrostatic boundary conditions imply that the mean microion density $\rho_{av}$ and electric field $\bm{E}$ vary with $z$. Since the fluctuating Maxwell stresses originate from the force density $\rho E$,  the excess noise is now multiplicative, with a $z$ dependence. The microions screen the potential so that the electric field decays away from the wall. Upto linear order in $\phi_0$, ignoring the colloid charge and the microion density fluctuations, the screened electric field is given by the linearized Poisson-Boltzmann equation solved with the above boundary conditions; $\bm{E}_{s} = \bm{\hat{z}} E_{s0}e^{-\kappa z}$, where $E_{s0} = \phi_{0}/ \kappa^{-1}$. For small colloid charge $Q$ and to lowest order in charge density fluctuations the force density in the Stokes equation is approximately $\rho^0 \bm{E}_s \Theta (z)$, where $\Theta$ is the Heaviside function and $\rho^0$ is the microion distribution in case I. The screened electric field recieves corrections due to $\rho^0 $ and image charges required to satisfy boundary conditions, but for large screening and small Q this effect is seen to be very small. The buoyant weight of the particle adds a force density $W\delta(\bm{r}-\bm{R}(t))$ to (\ref{momcons}).

We use the Green's function $\mathcal{G}(\bm{r},\bm{r}^\prime)$ for the Stokes equation with a no-slip no-penetration wall in the $xy$ plane as obtained by Blake \cite{blake} to solve (\ref{momcons}) for the velocity field $\bm{v}$, including contributions from equilibrium thermal fluctuations and fluctuating Maxwell stresses proportional to $\bm{E}$. Evaluating $\bm{v}$ at the colloid position $\bm{R}(t)$ gives the Langevin equation for the colloid in the form (\ref{colloid}), with an $\bm{R}$ dependent velocity $\bm{v}_0(\bm{R})$ with a zero corresponding to the minimum of Squires's effective potential. The thermal noise $\bm{F}$ and excess noise $\zeta$ are now multiplicative, with correlations
\begin{eqnarray}
\langle \zeta^\prime_i(t,\bm{R}(t))\zeta^\prime_j(t^\prime,\bm{R}(t^\prime)) 
\rangle &=& \int \mbox{d}^3 k  \frac{1}{\kappa^2 + k^2} 
\mbox{e}^{-(\kappa^2 + k^2) (t-t^\prime)}   \mathcal{H}_i(\bm{k},z(t), k_z) 
 \nonumber \\
&\times&\mathcal{H}_j(-\bm{k}_\perp,z(t^\prime), -k_z) ,
\label{nonst} \\
\langle F^\prime_i(t,\bm{R}(t)) F^\prime_j(t^\prime,\bm{R}(t^\prime)) 
\rangle &=& \delta(t-t^\prime) \int \mbox{d}^3k 
\mathcal{H}_i^{\prime}(\bm{k}_\perp,z(t),k)
\nonumber  \\
&\times& \mathcal{H}_j^{\prime}(-\bm{k}_\perp , z(t),-k) \mbox{d}\bm{p} ,
\end{eqnarray}
where the function $\mathcal{H}_i$ and $\mathcal{H}^\prime_{i}$ are the 
Fourier transforms of $\mathcal{G}_{zi}(\bm{r}_{\perp},z,z') \exp(-\kappa z')$ and 
$\mathcal{G}_{zi}(\bm{r}_{\perp},z,z')$ respectively with respect 
to $\bm{r}_{\perp}$ and $z'$.
By analogy to (7), we estimate the ratio of excess noise strength to the thermal noise, 
now as a function of $z$, by replacing $z(t)$ and $z(t^\prime)$ in (\ref{nonst}) by a fixed $z$.
\begin{eqnarray}
\mathcal{R}_{ij} = \frac{\mathcal{R}^{\zeta}(z)}{\mathcal{R}^{F}(z)} = 
\frac{\int_0 ^\infty \mbox{d}t\langle \zeta_i(0,z)\zeta_j(t,z) \rangle }
{\int_0 ^\infty \mbox{d}t\langle F_i(0,z) F_j(t,z) \rangle}
\end{eqnarray} 
The boundary conditions in the velocity field imply $\mathcal{R}^{F}(0) = 0 =
\mathcal{R}^{\zeta}(0)$; $\mathcal{R}^{F}$ rises to its saturation value within
$1-2$ particle radii, whereas $\mathcal{R}^{\zeta}$ peaks at $z = \kappa^{-1}$
and falls off due to exponential fall in the field strength with increase in
$z$ so that the tail penetrates more into the bulk for smaller $\kappa a$. The
peaks get sharper and the peak value larger with increase in $\kappa a$.  Since
$\mathcal{R}^{\zeta}$ and $\mathcal{R}^{F}$ are anisotropic to varying degree,
$\mathcal{R}_{ij}$ is anisotropic and $\mathcal{R}_{zz}/\mathcal{R}_{xx}$ is
$z$  dependent close to the wall saturating to a constant value close to $8$, a
few particle radii away from $z=0$. $\mathcal{R}_{ij}$ scales as $\epsilon
V_0^2 \kappa a a_{mic}/k_{B}T$; typically in experiments, $V_{0} \sim \mbox{a
few volts}$, $a= 0.1 \mu m$ to $10 \mu m$ and $\kappa a \sim 1$ so that the
ratio is at least $\mathcal{O}(1)$ within a few $a$ of $z=\kappa^{-1}$. The
ratio has dipped in magnitude compared to case I due to exponential fall in the
field magnitude and the wall boundary conditions.

As a result of the excess noise proportional to $\bm{E}$, the steady-state
probability distribution $P_\infty(\bm{R})$ is not given by a Boltzmann weight
determined by Squires's effective potential. To obtain $P_\infty$ we must solve
the equation for $\bm{R}$ treating carefully the multiplicative and coloured
noises in (6). The equation is driven by a continuous, weighted, linear
superposition of independent, multiplicative Ornstein--Uhlenbeck (OU) \cite{OU}
noise sources. We construct the Fokker Planck equation corresponding to (6) by
generalizing the results of \cite{rffoxfunc} for a single multiplicative OU
noise. Since the weights and relaxation times of the constituent noises in our
problem are finite as a result of Debye screening, all integrals arising in
this procedure are convergent. The resulting FP equation takes the form
\begin{eqnarray}
& & \hspace{-0.25in} \frac{\partial P}{\partial t} = -\frac{\partial(W P)}{\partial z} + \frac{\partial}{\partial z} \sqrt{\mathcal{R}_{zz}} \frac{\partial}{\partial z} ( \sqrt{\mathcal{R}_{zz}} P)
\nonumber \\ & &
\hspace{-0.1in}
+ \int \mbox{d} \tau \frac{\partial}{\partial z}[G_\tau \frac{\partial}{\partial z}(G_\tau +\tau G_\tau W^\prime -\tau G^\prime_\tau W )P ]
\label{fokpl1}
\end{eqnarray}
where we have schematically replaced integrals over wave number by integrals
over relaxation time $\tau$; $\tau$ goes as $[D( k^2 + \kappa^2)]^{-1}$ and
$G_\tau(R) =(k^2+\kappa^2)^{-1} \mathcal{H}(\bm{p},z,k) $. $P_{\infty}(R)$ is
obtained by solving (\ref{fokpl1}) for the zero flux condition and $U_{eff} =
-k_B T \ln P_{\infty}(R)$. The probability distribution $P_{\infty}^0$
analogous to that of \cite{squiresM}, containing only the bare thermal noise,
is obtained by setting $G_\tau = 0$ in (\ref{fokpl1}); the corresponding
effective potential $U_{eff}^0 \equiv -k_B T \ln P_{\infty}^0$. The two
potentials show large differences near the wall where the excess noise is
significant. $S_{R}=(U_{eff}^0)^{\prime \prime} \mid_{R_{av}}
/(U_{eff})^{\prime \prime}\mid_{R_{av}} $, gives the relative confining
strengths of the two potentials and $S_R>>1$ for $R_{av}$ close to the wall
($S_R = 2.7$ for Fig.2.); i.e. $U_{eff}$ is shallower than $U_{eff}^0$. The
excess noise causes the colloid to explore a wider range of $z$.  This
nonequilibrium effect is strongest if the colloid is heavy, so that its mean
position is close to the wall. Fig. 2 shows the comparison between $U_{eff}$
and $U_{eff}^0$ for a colloid with mass density corresponding to (a) silica and (b)
iron.  

Lastly, let us check that the neglect of the advection of microions by the hydrodynamic velocity field $\bm{v}$ was not too bad. The appropriate Peclet number Pe$= v/D \kappa^2 a$, the ratio of the rate at which the colloid shears the medium to the rate at which microion densities relax. The Langevin equation for the colloid implies a typical deviation from mean position and hence a typical value for the right-hand-side of (6). Using this, with a surface charge density 
$\simeq 10^{-4}-10^{-3}$ C m$^{-2}$ which is a reasonable approximation for silica particles 
\cite{silica}, we find Pe$= 0.35$, $0.12$, for $\kappa a = 3$, $5$ 
respectively. Thus the neglect of advection is not a bad approximation. 

In summary, we have formulated the statistical dynamics of a single colloidal
particle in a static electric field. We show that thermal agitation of
counterion and impurity charge densities, in the presence of the imposed field,
leads to noisy Maxwell stresses and hence to an additional noise term,
proportional to the field, in the effective Langevin equation for the colloidal
particle. This noise is nonequilibrium in nature and highly anisotropic, and
leads to strong, frequency-dependent departures from the
Fluctuation-Dissipation Theorem. The effective potential as inferred \cite{squiresM} 
by combining Stokes drag with the mean velocity when the particle is displaced from 
its steady-state position
differs substantially from that obtained by taking the logarithm of the
steady-state probability distribution. Our results are
quantitatively testable, e.g., in experiments such as those of \cite{Larsen},
and should form an essential ingredient in understanding the nonequilibrium
steady states of colloids in electric fields. Generalizations of our treatment 
to include oscillatory fields, as well as more than one colloidal particle, 
are underway.



\end{document}